\documentclass[preprint]{ptephy_v1}

\preprintnumber{XXXX-XXXX} 
\usepackage{hyperref}
\usepackage{multirow}
\usepackage{color}

\begin{document}

\title{First Study of the PIKACHU Project: Development and Evaluation of High-Purity Gd$_3$Ga$_3$Al$_2$O$_{12}$:Ce Crystals for $^{160}$Gd Double Beta Decay Search}

\author[1]{Takumi Omori }
\affil[1]{Graduate School of Science and Technology, University of Tsukuba,  Tsukuba, Ibaraki, 305-8571, Japan }

\author[2,*]{Takashi Iida}
\affil[2]{Institute of Pure and Applied Sciences, University of Tsukuba,  Tsukuba, Ibaraki, 305-8571, Japan
\email{tiida@hep.px.tsukuba.ac.jp}}

\author[3]{Azusa Gando}
\affil[3]{Research Center for Neutrino Science, Tohoku University, Sendai 980-8578, Japan}

\author[4]{Keishi Hosokawa}
\affil[4]{Kamioka Observatory, Institute for Cosmic Ray Research, University of Tokyo, Kamioka, Gifu 506-1205, Japan}

\author[5,6]{Kei Kamada}
\affil[5]{New Industry Creation Hatchery Center, Tohoku University, 6-6-10 Aramaki Aza Aoba, Aoba-ku, Sendai, Miyagi, 980-8579, Japan}
\affil[6]{C\&A Corporation, 1-16-23 Ichibancho, Aoba-ku, Sendai, Miyagi, 980-0811, Japan}

\author[7]{Keita Mizukoshi}
\affil[7]{Institute of Space and Astronautical Science, Japan Aerospace Exploration Agency, Sagamihara, Kanagawa, 252-5210, Japan}

\author[6]{Yasuhiro Shoji} 

\author[5,6]{Masao Yoshino}

\author[8]{Ken-Ichi Fushimi}
\affil[8]{Division of Science and Technology, Tokushima University, 2-1 Minami Josanjima-cho Tokushima city, 
Tokushima 770-8506, Japan}


\author[1]{Hisanori Suzuki}
\author[1]{Kotaro Takahashi}


\begin{abstract}%
Uncovering neutrinoless double beta decay (0$\nu$2$\beta$) is crucial for confirming neutrinos' Majorana characteristics. The decay rate of 0$\nu\beta\beta$ is theoretically uncertain, influenced by nuclear matrix elements that vary across nuclides. To reduce this uncertainty, precise measurement of the half-life of neutrino-emitting double beta decay (2$\nu$2$\beta$) in different nuclides is essential. 

We have launched the PIKACHU (Pure Inorganic scintillator experiment in KAmioka for CHallenging Underground sciences) project to fabricate high-purity Ce-doped Gd$_{3}$Ga$_{3}$Al$_{2}$O$_{12}$  (GAGG) single crystals and use them to study the double beta decay of $^{160}$Gd. Predictions from two theoretical models on nuclear matrix element calculations for 2$\nu$2$\beta$ in $^{160}$Gd show a significant discrepancy in estimated half-lives, differing by approximately an order of magnitude. If the lower half-life estimation holds true, detecting 2$\nu$2$\beta$ in $^{160}$Gd could be achievable with a sensitivity enhancement slightly more than an order of magnitude compared to prior investigations using Ce-doped Gd$_2$SiO$_5$ (GSO) crystal. We have successfully developed GAGG crystals with purity levels surpassing previous standards through refined purification and selection of raw materials. Our experiments with these crystals indicate the feasibility of reaching sensitivities exceeding those of earlier studies. This paper discusses the ongoing development and scintillator performance evaluation of High-purity GAGG crystals, along with the anticipated future prospects of the PIKACHU experiment.

\end{abstract}

\subjectindex{xxxx, xxx}

\maketitle

\section{Introduction}
The neutrino-less double beta (0$\nu$2$\beta$) decay is a significant physical process beyond the Standard Model, offering definitive proof of the Majorana nature of neutrinos which are their own antiparticles. The half-life of 0$\nu$2$\beta$ ($T_{1/2}^{0\nu}$) correlates inversely with the square of the effective electron neutrino mass ($m_{\beta\beta}$), shedding light on the neutrino's absolute mass. 
While 0$\nu$2$\beta$ remains undetected, the KamLAND-Zen experiment currently leads in sensitivity for this decay, with the lower limit of half-life, 2.3 $\times$ 10$^{26}$ years for $^{136}$Xe \cite{Kamland}. Conversely, double beta decay with neutrino emission (2$\nu$2$\beta$), as predicted by the Standard Model, has been observed in several nuclei \cite{2nbb}. Numerous research groups continue to investigate various nuclei, as determining the half-life of 2$\nu$2$\beta$ can help refine \textcolor{black}{the} accuracy of the nuclear matrix elements \textcolor{black}{(NME)} \cite{NME}.

The PIKACHU project, a Pure Inorganic scintillator experiment in KAmioka for CHallenging Underground sciences, focuses on double beta (2$\beta$) decay of $^{160}$Gd. This isotope, a potential 2$\beta$ decay candidate, is characterized by a high natural abundance (21.9$\%$) and a low $Q$-value (1.730~MeV \cite{Qval}) compared to other candidate nuclei. 
This enables the use of a substantial amount of target nuclei in experiments. However, it is crucial to minimize the background noise from environmental radioactivity, especially radiation close to the $Q$-value of $^{160}$Gd. Currently, the most sensitive $^{160}$Gd experiment has been conducted in Ukraine using a 2-inch square Ce-doped Gd$_{2}$SiO$_{5}$ (GSO) crystal containing 103.6 g of $^{160}$Gd \cite{Ukraine}. This experiment obtained a lower limit of 2.3 $\times$ 10$^{21}$ years for 0$\nu$2$\beta$ and 1.9 $\times$ 10$^{19}$ years for 2$\nu$2$\beta$.

The sensitivity of the previous study was limited by background noise from Uranium or Thorium decay series \textcolor{black}{(U/Th)}, inherent contaminants in their GSO crystal \cite{Ukraine}. Our research utilizes Ce-doped Gd$_{3}$Ga$_{3}$Al$_{2}$O$_{12}$ (GAGG) scintillator crystals for the 2$\beta$ decay investigation. GAGG is a type of inorganic scintillator for which large single crystal growth up to 4 inches in diameter via the Czochralski method has been well-established \cite{Kochurikhin}. For our experiments, we use a GAGG crystal measuring 6.5 cm in diameter and 14.5 cm in length, containing 710 g of $^{160}$Gd. GAGG's superior light output ($\sim$ 50,000 photons/MeV) compared to GSO's ($\sim$ 10,000 photons/MeV) and its pulse shape distinguishing capability (PSD) for particle identification \cite{Tamagawa} provides us with an edge in 2$\beta$ decay searches. Nevertheless, developing low radioactivity GAGG (High-purity GAGG) remains a critical research task.

Predictions for the half-life of 2$\beta$ decay are variable due to uncertainties in matrix element calculations. For $^{160}$Gd's 2$\nu$2$\beta$ decay, two primary theoretical half-lives have been predicted. The theory that calculated the NME using the pseudo-SU(3) model predicts a half-life of 6.02 $\times$ 10$^{21}$ years \cite{Hirsch}, while the NME calculated with the quasiparticle random-phase approximation (QRPA) method \cite{Hinohara} is larger than that in Ref.~\cite{Hirsch} and predicts a half-life of 4.7 $\times$ 10$^{20}$ years if the same phase space factor is used.
If the experiment can be performed with sensitivity about an order of magnitude higher than the previous experiment, the search sensitivity will approach the theoretical prediction in Ref. \cite{Hinohara}, and we can expect to find 2$\nu$2$\beta$ of $^{160}$Gd, or to constrain the theoretical model by updating the lower half-life limit.

\textcolor{black}{The PIKACHU project aims to discover $^{160}$Gd 2$\nu$2$\beta$ by proceeding with the following sequence of steps.}

\begin{enumerate}
    \item \textcolor{black}{We will develop large-size and High-purity GAGG crystals. In the first step, we aim to produce high-purity crystals that are one order of magnitude cleaner than current crystals, and in the next step, we aim to produce ultra-high purity crystals that are two or more orders of magnitude higher in purity.}

    \item \textcolor{black}{Using GAGG with an order of magnitude higher purity, measurements with sensitivity exceeding that of previous studies will be performed. (PIKACHU experiment Phase 1)}

    \item \textcolor{black}{Long-term measurements with a sensitivity about one order of magnitude higher than the previous study will be performed using several ultra-high purity GAGG crystals with a purity two orders of magnitude higher than the conventional GAGG crystals. (PIKACHU experiment phase 2)}

    \item \textcolor{black}{The data obtained will be analyzed with the aim of discovering 2$\nu$2$\beta$ of $^{160}$Gd. If no discovery is made, the lower half-life limit will be updated to provide feedback on the theory of NME calculations.}

\end{enumerate}

\textcolor{black}{This paper discusses the development of High-purity GAGG for the PIKACHU experiment Phase 1, its performance assessment, background contamination estimation, and the evaluated search sensitivity for 0$\nu$2$\beta$.}


\section{Conventional GAGG single crystal}
\begin{figure}[h]
\centering\includegraphics[width=13cm]{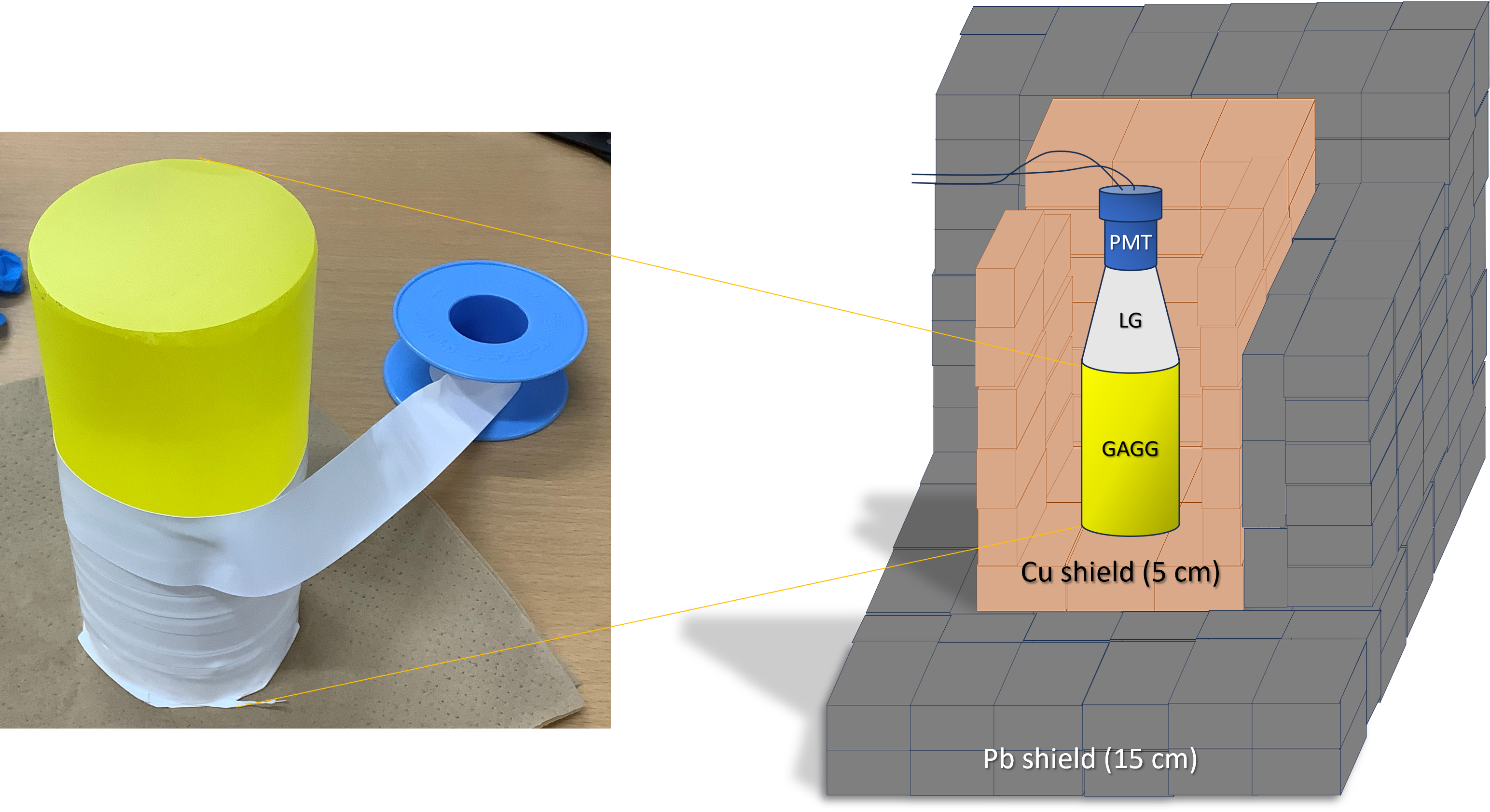}
\caption{(Left) Conventional GAGG crystals used in the background study of the PIKACHU experiment, wrapped with 200 $\mu$m thick Teflon tape as reflective material.
(Right) Schematic of the GAGG background level measurement setup at Kamioka.}
\label{Fig1}
\end{figure}

In the preliminary stage of the \textcolor{black}{High-purity GAGG development}, standard commercial GAGG single crystals were used for background investigation.
The crystal used was 6.5 cm in diameter and 14.5 cm in length single crystal (4N GAGG crystal) shown in Figure \ref{Fig1} left, whose time constant was shortened by co-doping Mg to the main raw material of 4N-level purity \cite{canda1}.
To construct a detector, these 4N GAGG crystals were combined with a 2-inch photomultiplier tube (PMT) with a bialkali photo-cathode and an acrylic light guide enhancing the efficiency of light collection. We then set this detector within a radiation shield (15 cm of lead and 5 cm of copper) situated in the KamLAND experimental area in Kamioka, Gifu Prefecture, Japan. 
This setup was intended to assess the background levels of the crystal under a low-background environment, as illustrated in Figure \ref{Fig1} right. 
Comprehensive details about the experiments and their analyses are detailed in the subsequent section on High-purity GAGG crystals.

Our measurements conducted in Kamioka revealed that the background level in the $Q$-value region was approximately ten times higher than that observed with the previously used GSO crystals in Ukrainian studies \cite{Ukraine}. We deduced that a reduction in background levels by at least an order of magnitude would be imperative to achieve a sensitivity surpassing that of the earlier studies. This realization led us to the decision of developing High-purity GAGG crystals specifically for the PIKACHU experiment.

\section{Development of High-purity GAGG single crystal}

\subsection{Purification and measurement of materials of GAGG crystal}
Initial background measurements at Kamioka revealed that conventional 4N GAGG crystals had substantial levels of U/Th chain radioactive impurities. This necessitated a reassessment of both the raw materials and the crystal growth process. The primary constituents of GAGG crystals are gadolinium oxide (Gd$_2$O$_3$), gallium oxide (Ga$_2$O$_3$), and aluminum oxide (Al$_2$O$_3$), blended in an approximate weight ratio of (Gd$_2$O$_3$:Ga$_2$O$_3$:Al$_2$O$_3$)=(1:0.5:0.2). A minor addition of cerium oxide (CeO$_2$) is also included. The initial step involved assessing the main raw materials for radioactive impurities using High-Purity Germanium (HP-Ge) detectors, specifically those employed in the XMASS \cite {XMASS} \textcolor{black}{or} SK-Gd \cite{SK-Gd} experiment at the Kamioka Underground Observatory.

The evaluation indicated exceptionally high upstream impurities in the $^{238}$U  series in Gd$_2$O$_3$, quantified at 1750 $\pm$ 221 mBq/kg, and significant levels in the $^{232}$Th series.
Al$_2$O$_3$ also yielded high background levels of $^{238}$U and $^{232}$Th in the HP-Ge detector, indicating the need to reduce impurities in the raw material.
On the other hand, Ga$_2$O$_3$ and CeO$_2$ were found to have impurity levels below the detection limit, leading to the decision to use existing supplies of these materials for subsequent crystal growth.

Purification of Gd$_2$O$_3$ was executed by the same supplier responsible for the high-purity gadolinium sulfate in the SK-Gd experiment. The process involved dissolving Gd$_2$O$_3$ powder into acid to selectively remove U/Th from the solution, similar to the method used in the SK-Gd experiment. The detailed purification process is described in the SK-Gd paper \cite{SK-Gd}. As Al$_2$O$_3$ is acid-insoluble, purifying it through the same method was not feasible. Therefore, we procured four different Al$_2$O$_3$ samples from three vendors and measured their impurity levels using an HP-Ge detector. The sample with the lowest impurity content was selected for crystal growth.

Table \ref{Raw material} \textcolor{black}{summarizes} the HP-Ge measurement results of the impurity content in the raw materials used for both 4N and High-purity GAGG crystals, covering all four main raw materials. For $^{235}$U and $^{40}$K, impurity content was directly calculated from the gamma-ray intensities of their emissions. While for $^{238}$U, since the branching ratio of gamma-ray emission is small and difficult to detect, a weighted average was taken from the values of $^{234}$Th and $^{234}$Pa, assuming radioactive equilibrium. \textcolor{black}{The weights were calculated using the gamma-ray branching ratio, the detection efficiency obtained from simulations, and the statistical error.}
The gamma-ray intensity of $^{228}$Ra was used to calculate the impurity level for $^{232}$Th.

\begin{table}[h]
\caption{Details of the radioactive impurity levels and 90$\%$ upper limits in Gd$_{2}$O$_{3}$, Al$_{2}$O$_{3}$, Ga$_{2}$O$_{3}$ and CeO$_2$ in mBq/kg for both High-purity GAGG and 4N GAGG. The raw materials labeled as pure and 4N represent those used for High-purity GAGG and 4N GAGG, respectively. The Ga$_{2}$O$_{3}$ initially used for 4N GAGG was of high purity, so the same Ga$_{2}$O$_{3}$ and CeO$_2$ were used for the High-purity GAGG crystals.}
\label{Raw material}
\centering
\small
\begin{tabular}{lcccccc}
\hline\hline
\multirow{2}{*}{Radioactivity}&\multicolumn{2}{c}{Gd$_{2}$O$_{3}$}&\multicolumn{2}{c}{Al$_{2}$O$_{3}$}&\multirow{2}{*}{Ga$_{2}$O$_{3}$} &\multirow{2}{*}{CeO$_2$}\\
&\textcolor{black}{high-purity}&4N&\textcolor{black}{high-purity}&4N&& \\\hline
$^{238}$U ($^{234}$Th \& $^{234}$U equiv.)& $<$ 16.30 & 1750 $\pm$ 221 & $<$ 28.26 & 476.0 $\pm$ 43.5 & $<$ 69.2 & $<$ 59.0 \\
$^{235}$U& $<$ 10.03 & 130 $\pm$ 40 & $<$ 7.82 & $<$ 21.08 & $<$ 10.2 & $<$ 15.5\\
$^{232}$Th ($^{228}$Ra equiv.)& $<$ 0.96 & 270 $\pm$ 12 & 5.85 $\pm$ 2.80 & 15.95 $\pm$ 6.58 & $<$ 10.8 & 4.4 $\pm$ 1.9\\
$^{40}$K& $<$ 2.70 & 84.8 $\pm$ 28.7 & $<$ 36.58 & $<$ 96.48 & $<$ 35.8 & $<$ 23.5\\
\hline\hline
\end{tabular}
\end{table}

\subsection{Crystal growth of High-purity GAGG crystal}
The process of growing High-purity GAGG crystals for the PIKACHU experiment employs the Czochralski (Cz) technique, which is a standard method in the industry and uses a radio frequency heating system. This technique is also used for commercially available GAGG single crystals. Detailed schematics of the Cz growth apparatus are available in Ref. \cite{Kochurikhin}, \cite{canda1}. High-purity GAGG crystal growth was performed using commercially available 4N purity CeO$_{2}$, $\beta$-Ga$_{2}$O$_{3}$, selected $\alpha$-Al$_{2}$O$_{3}$ and purified Gd$_{2}$O$_{3}$ powders. Nominally, starting powders were mixed according to the formula of (Ce$_{0.01}$, Gd$_{0.99}$)$_{3}$Ga$_{3}$Al$_{2}$O$_{12}$. Typical growth rates were 0.9 mm/min and the crystal diameter was around 55 mm. Crystals were grown from a 110 mm diameter Ir crucible under the N$_{2}$ with 30$\%$ CO$_{2}$ atmosphere. The $\langle$100$\rangle$ oriented GAGG single crystal was used as the seed. The grown crystal was cut into 52 mm length and polished into cylinders 54 mm $\phi$ in diameter as a prototype. \textcolor{black}{The surface of the crystal was then chemically polished with a mixture of phosphoric acid and sulfuric acid.}

\begin{figure}[h]
\centering\includegraphics[width=4.5in]{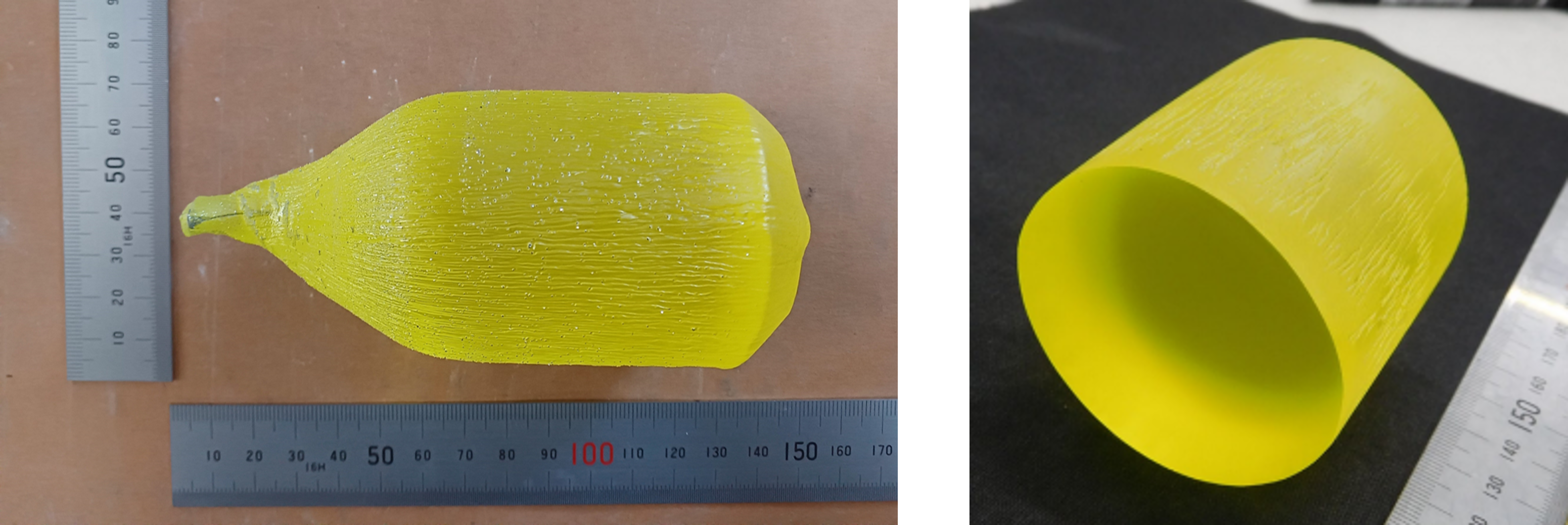}
\caption{(Left) the High-purity GAGG single crystal before being cut. The left end of this crystal indicates the seed crystal side.  (Right) High-purity GAGG after cutting. The crystal measured 54 mm in diameter and 52 mm in length, with a weight of 784 g.}
\label{Crystal1}
\end{figure}

\section{Performance evaluation of High-purity GAGG in Kamioka}
To effectively search for double beta decay using GAGG, it is crucial to assess its energy resolution, PSD, and inherent radioactive background. This section details the performance evaluation of High-purity GAGG, with the testing carried out at the Kamioka underground observatory. The detector setup for these evaluations comprised High-purity GAGG (as illustrated in Fig. \ref{Crystal1}), an acrylic light guide, and a PMT. These components were connected using an optical grease (TSK5353) that boasts high light transmittance, and the entire assembly was wrapped in reflective material and light-shielding tape. This configuration ensures that light emitted from the GAGG is channeled through the light guide and subsequently detected by the PMT. The PMT model utilized was the R6231-100 from Hamamatsu Photonics, featuring a \textcolor{black}{super-}bialkali photocathode.
This PMT is particularly suited for this application due to its sensitivity in the wavelength range of about 300 to 650 nm, aligning well with the GAGG emission wavelength of approximately 520 nm. The PMT was operated at a voltage of 1200 V. Signals from the PMT were \textcolor{black}{input directly to} a digitizer, specifically the DT5720 model from CAEN\textcolor{black}{, without amplifier to prevent saturation in high energy region.} DT5720 is equipped with a 12-bit ADC and a sampling rate of 250 MS/s. Analog signals from the PMT were sampled every 4 nsec, covering a range of 4.096 $\mu$sec up to 1024 channels.

\subsection{Energy calibration and Pulse Shape Discrimination}
\textcolor{black}{First, a $\gamma$-ray source was irradiated to High-purity GAGG to investigate the light yield and energy resolution. The gain curve of PMT \cite{r6231} was used to obtain the gain value at an applied voltage of 1200 V, and the output charge of a single-photoelectron event was calculated. Next, the detected charge was obtained from the waveform of the data for the $^{137}$Cs event, and the number of detected photo-electrons (p.e.) was calculated from the ratio of the detected charge of $^{137}$Cs to that of a single photoelectron, yielding a value of 9,730 p.e./MeV. The quantum efficiency of the super-bialkali photocathode is about 15\% at 520 nm, the emission wavelength of GAGG, and the emission of High-purity GAGG was calculated to be 64,730 photons/MeV. This value is roughly consistent with the typical light yield of GAGG.}

\begin{figure}[h]
\centering\includegraphics[width=5in]{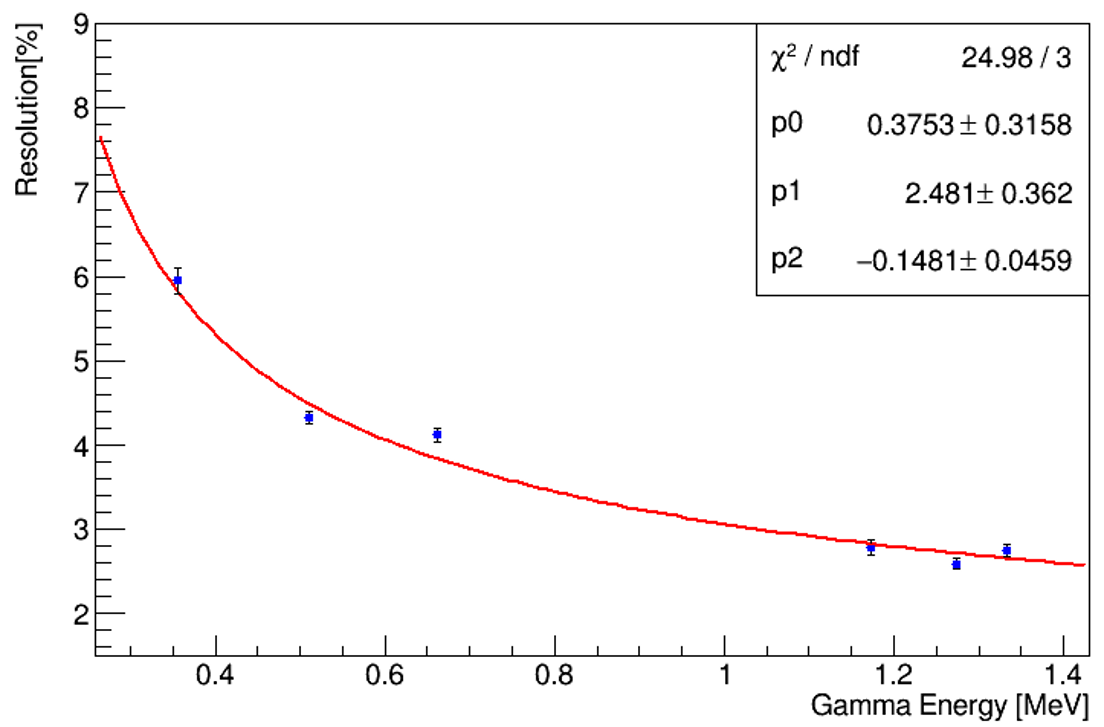}
\caption{Energy resolution characteristics of High-purity GAGG crystal concerning $\gamma$-ray energy. Red solid line represents the fitting result by function shown in the text.}
\label{resolution}
\end{figure}

The energy calibration of High-purity GAGG was achieved by exposing the crystal to $\gamma$-rays from various radioactive sources ($^{137}$Cs, $^{60}$Co, $^{22}$Na, and $^{133}$Ba) and linearly fitting the correlation between ADC values and the $\gamma$-ray energies.
Additionally, the energy resolution ($\sigma$) of GAGG was also evaluated from the energy spectrum of them as follows; 5.95 $\pm$ 0.15$\%$, 4.32 $\pm$ 0.07$\%$, 4.12 $\pm$ 0.08$\%$, 2.78 $\pm$ 0.09$\%$, 2.59 $\pm$ 0.07$\%$ and 2.74 $\pm$ 0.07$\%$ at energies of 356, 511, 662, 1173, 1274 and 1333 keV respectively. 
Figure \ref{resolution} displays the $\gamma$-ray energy dependence of the resolution for High-purity GAGG, with the red solid line representing the fit of the data to the equation $y = \mathrm{p_0} + \frac{\mathrm{p_1}}{\sqrt{x+\mathrm{p_2}}}$. By extrapolating this energy dependence to the $Q$-value of $^{160}$Gd (1.730~MeV), the energy resolution at the $Q$-value was estimated to be 2.35\textcolor{black}{ $\pm$ 0.07}$\%$.

For the PIKACHU experiment, a high-quality PSD performance between $\alpha$ and $\beta$ ($\gamma$)  radiations is crucial \textcolor{black}{to remove the $\alpha$-ray background analytically}. To evaluate this, we conducted an 18-hour (66,338 sec) background measurement in the shield (as shown in Fig. \ref{Fig1}), recording 460,266  events. We defined a ``Ratio'' to distinguish signal waveforms, calculated as the ratio of the integral value for 200 nsec from the onset to that of the entire waveform for each signal.

\begin{figure}[h]
\centering\includegraphics[width=5in]{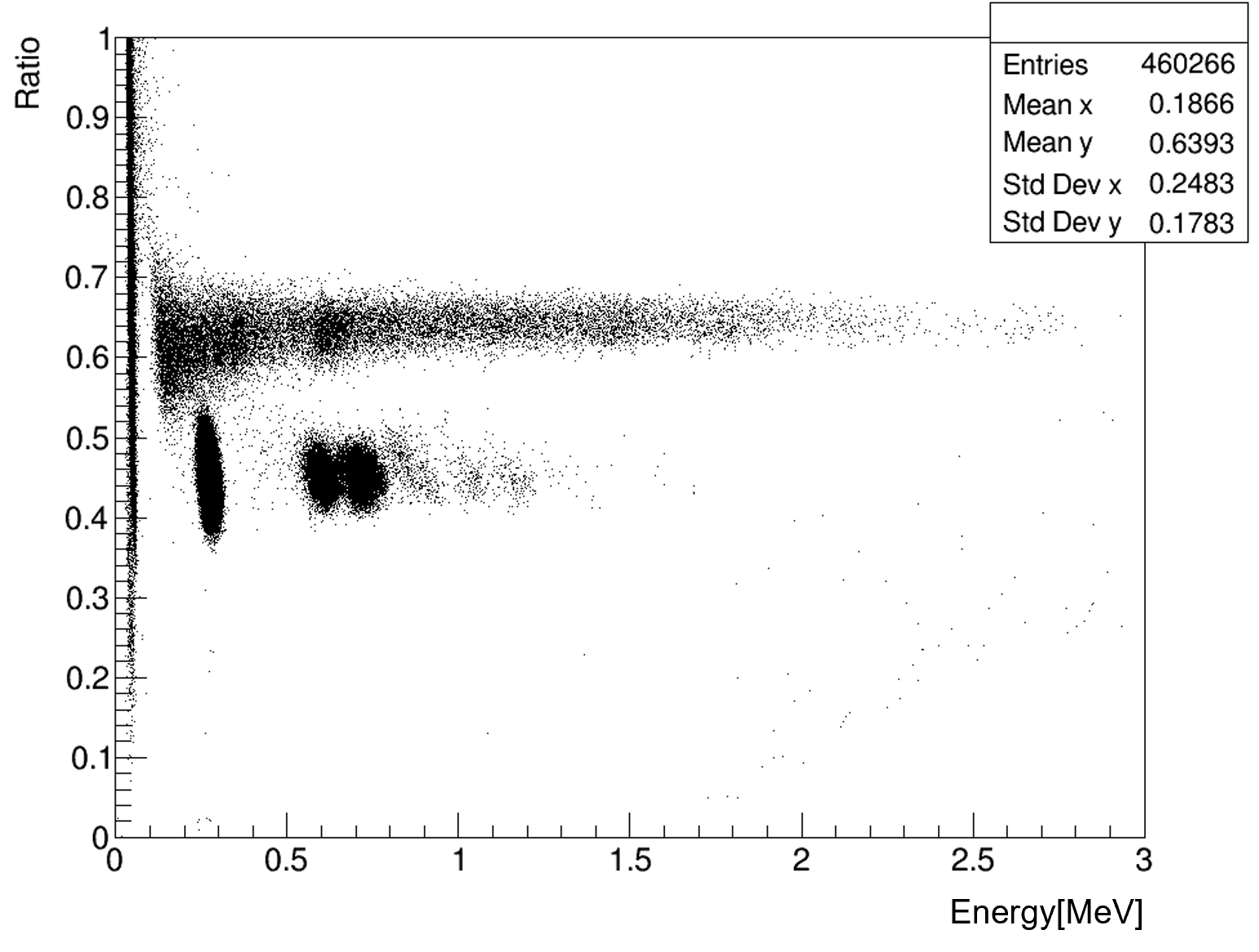}
\caption{2D distribution of Ratio versus energy. The upper event band is the $\beta$ ($\gamma$)-ray events, and the lower event band is the $\alpha$-ray events.}
\label{Ratio map}
\end{figure}

Figure \ref{Ratio map} presents a 2-dimensional distribution of event-by-event Ratio versus energy. It is evident that events with energies $>$ 0.15 MeV are distinctly separated at a Ratio of 0.55. Since $\alpha$-ray signals attenuate slower than $\beta$ ($\gamma$)-ray signals, events with Ratio $<$ 0.55 are identified as $\alpha$-rays, while those with higher Ratios are $\beta$ ($\gamma$)-rays in Fig. \ref{Ratio map}. The $\alpha$-ray events plotted around 0.2 to 0.3 MeV are attributed to $^{152}$Gd $\alpha$ decay in GAGG, and plots with energy $<$ 0.15 MeV are considered noise, thus excluded from the analysis. Events with Ratio $<$ 0.4 and energy $>$ 1.5 MeV are likely due to sequential decays within one sampling window, caused by nuclides in natural decay series; for example $^{212}$Bi ($Q_{\beta} =$ 2.25 MeV, $T_{1/2} =$ 60.6 $\rm min$) $\rightarrow$ $^{212}$Po ($Q_{\alpha} =$ 8.78 MeV, $T_{1/2} =$ 299 $\rm nsec$) $\rightarrow$ $^{208}$Pb in $^{232}$Th decay series, or $^{214}$Bi ($Q_{\beta} =$ 3.27 MeV, $T_{1/2} =$ 19.9 $\rm min$) $\rightarrow$ $^{214}$Po ($Q_{\alpha} =$ 7.69 MeV, $T_{1/2} =$ 164.3 $\rm \mu sec$) $\rightarrow$ $^{210}$Pb in $^{238}$U decay series. These decay series exist as intrinsic contamination in GAGG.

\begin{figure}[h]
\centering\includegraphics[width=5in]{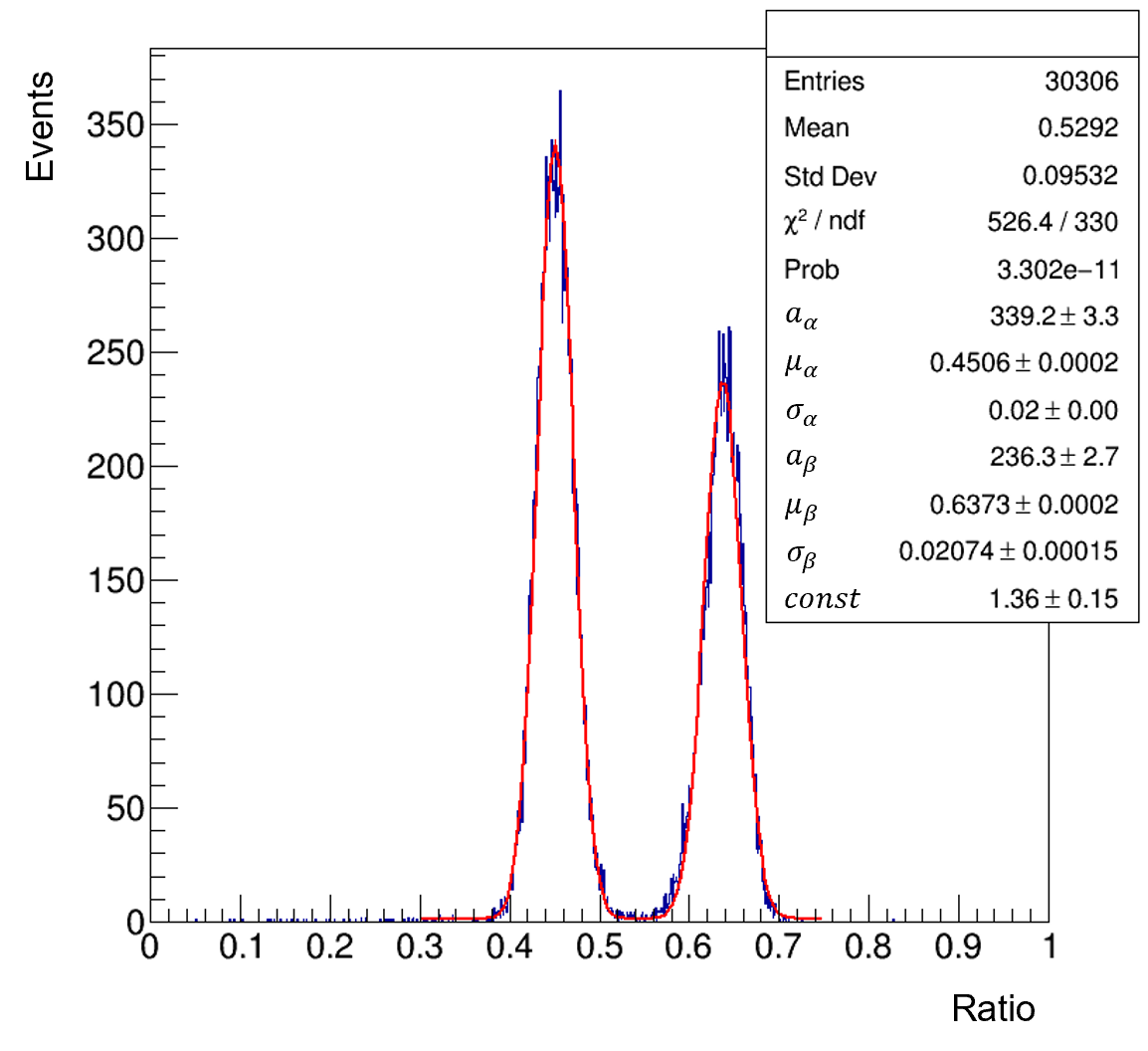}
\caption{Ratio distribution for events with energies 0.15 MeV to 3.0 MeV. The right peak is by $\beta$ ($\gamma$)-ray events and the left peak is by $\alpha$-ray events. The result of fitting two peaks with double Gaussian is represented by the red solid line. $\mu_{\alpha}$($\mu_{\beta}$) and $\sigma_{\alpha}$($\alpha_{\beta}$) represent the mean and sigma obtained by fitting to left (right) peak with Gaussian.}
\label{Ratio dst}
\end{figure}

Figure \ref{Ratio map}  demonstrates sufficient separation to analytically remove the $\alpha$-ray background. To quantify the PSD, a Figure of Merit (FoM) was used, indicating the effectiveness of the separation. FoM is calculated using Eq.~(\ref{1.1}), with $\mu_{\alpha}$ ($\mu_{\beta}$) and $\sigma_{\alpha}$ ($\sigma_{\beta}$) being fitting parameters as shown in Fig.~\ref{Ratio dst}. 
\begin{equation}\label{1.1}
\begin{split}
\mathrm{FoM} = \frac{|\mu_{\alpha}-\mu_{\beta}|}{\sigma_{\alpha}+\sigma_{\beta}}
\end{split}
\end{equation}
This figure displays the Ratio distribution for events with energies between 0.15 MeV and 1.5 MeV, fitted by a double Gaussian distribution. Substituting the fit results into Eq.~(\ref{1.1}), the FoM for High-purity GAGG was calculated to be $4.58\;\pm\;0.11$. In contrast, the FoM for 4N GAGG, derived from a similar analysis, was $2.30\;\pm\;0.02$. This result indicates that the PSD of High-purity GAGG is 1.99 times better than that of 4N GAGG, likely due to the higher luminescence of High-purity GAGG compared to \textcolor{black}{Mg co-doped} GAGG crystals.

\subsection{Background-level comparison}
\begin{figure}[h]
\centering\includegraphics[width=5.0in]{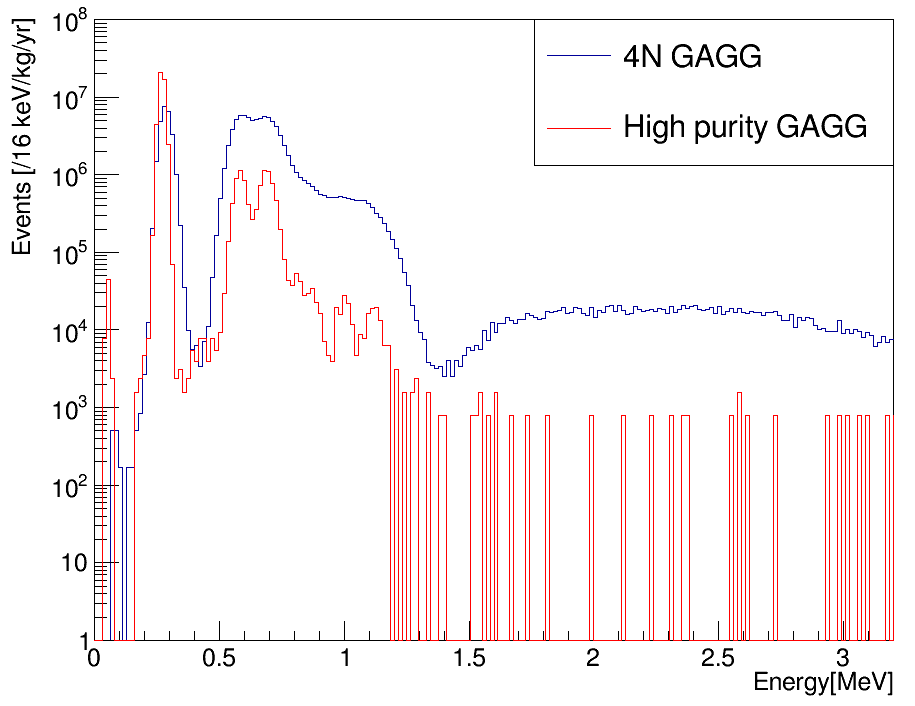}
\caption{Scaled $\alpha$-ray spectrum from impurity of each GAGG.}
\label{alpha spectrum}
\end{figure}

Due to the significant geometric separation between the GAGG crystal and the PMT in the detector setup, we can attribute $\alpha$-ray events predominantly to the $\alpha$ decay nuclides within the natural Uranium-Thorium decay series present in GAGG. Consequently, the $\alpha$-ray background spectrum becomes a vital indicator for estimating the internal impurities in GAGG. 
The ability to distinguish $\alpha$ and $\beta$($\gamma$)-rays through PSD not only facilitates the reduction of background but also enables more precise estimation of background by extracting alpha particles only from the data for comparison with simulations. This is an advantage unique to GAGG and is not present in the previous GSO studies.
Figure~\ref{alpha spectrum} illustrates the $\alpha$-ray background spectrum, derived using a straightforward criterion of Ratio $>$ 0.55. This figure compares the spectra of 4N GAGG (shown in blue) and High-purity GAGG (shown in red), with the vertical axis normalized for measurement time and the weight of the GAGG crystal.

It is obvious from Fig. \ref{alpha spectrum} that the amount of impurity in High-purity GAGG is reduced \textcolor{black}{compared to} that in 4N GAGG \textcolor{black}{as a whole}. Quantitatively, the $\alpha$-ray background rates are calculated as 4.24 $\times$ 10$^{{\color{black}7}}$ events/year/kg for High-purity GAGG and 1.15 $\times$ 10$^{8}$ events/year/kg for 4N GAGG. However, identifying specific impurities and conducting a more quantitative assessment of impurity content requires simulating the experimental background. Such simulations are instrumental in accurately replicating and analyzing the experimental conditions and results, thereby enabling a more detailed understanding of the impurity profile within the GAGG crystals.

\subsection{Detailed background study using Geant4-based simulation}

\begin{figure}[h]
\centering\includegraphics[width=5.0in]{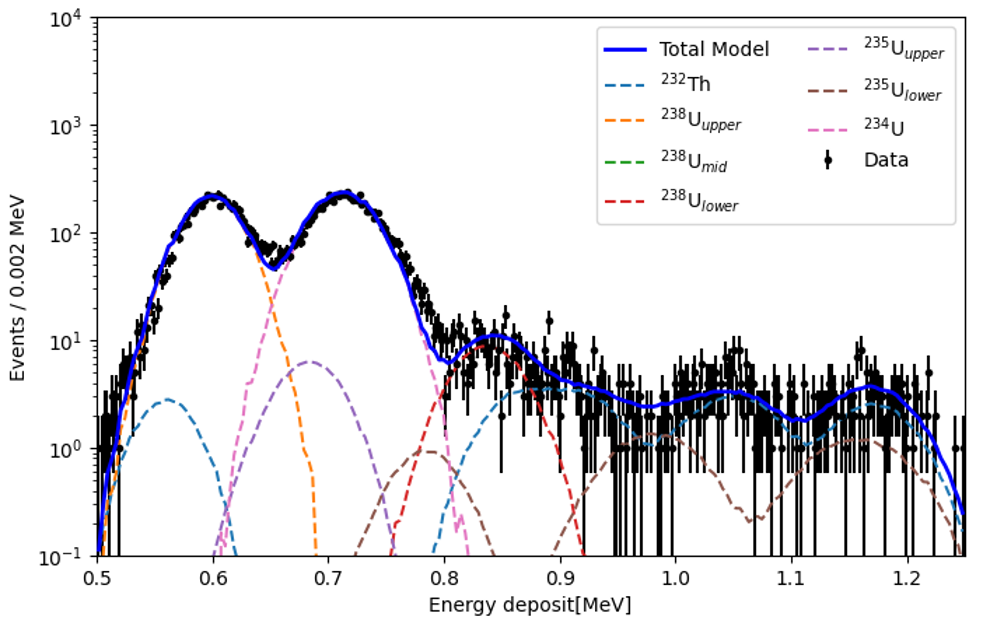}
\caption{
Result of fitting $\alpha$-ray spectra of High-purity GAGG with background model. In the graphical representation, the black points with error bars signify the measured $\alpha$-ray background. These measured data are then fitted using a background model, represented by the blue solid line. This background model is a composite of the spectra from various radioactive equilibrium groups, each of which is depicted as a dotted line in the graph. These groups are detailed in the graph's legend.
}
\label{alpha fit}
\end{figure}

 To enhance the understanding of background radiation due to radioactive impurities in GAGG, we employed Geant4-based simulations. The GAGG crystal model in Geant4 was cylindrical, with specified dimensions of 65 mm in diameter, 145 mm in length, and weighing 3.19 kg. The simulation incorporated impurities as primary particles from all nuclides in the $^{238}$U, $^{235}$U, and $^{232}$Th decay series, assumed to be uniformly distributed throughout the crystal's volume.  For the simulation of  $\alpha$-ray, $\beta^{\pm}$-ray and $\gamma$-ray energy deposits in the GAGG crystal, we used the G4DecayPhysics, G4RadioactiveDecayPhysics, and G4EmStandardPhysics modules from the GEANT4 \textcolor{black}{(v 11.0.2)} Reference PhysicsList library (Shielding). 
 These modules account for standard nuclear decay, spontaneous nuclear decay physics, and electromagnetic field interactions, covering $\alpha$ and $\beta$ decays as well as $\gamma$ rays from nuclear de-excitation. 
The simulation extracts the energy losses of radiation occurring between each specified nuclide.
We generated the energy deposit in the crystal using the GEANT4-based simulation without detector response e.g., energy resolution, thus we applied the actual energy resolution, as depicted in Fig. \ref{resolution}, to the energy deposit data. In scintillators, $\alpha$-rays exhibit a lower light yield per MeV of absorbed energy compared to $\gamma$-rays, a phenomenon commonly termed as quenching. This study quantifies detected radiation energy in electron-equivalent units (MeVee), resulting in an underestimation of the energy for $\alpha$-ray events compared to their actual energy levels. Conversely, this quenching effect was applied with necessitating the correction of the differences between real and simulated energy. 
We determined quenching coefficients by fitting the correlation between observed and actual energies with a quadratic function, and then adjusted the simulated $\alpha$-ray energy data accordingly. The $\alpha$-ray background spectrum (Fig. \ref{alpha spectrum}) was fitted with Probability Density Functions generated for each radioactive equilibrium. The $^{238}$U series was classified into several groups based on parent nuclides, and the $^{235}$U series was similarly categorized. The groupings are as follows : $^{238}$U series can be classified into $^{238}$U$_\mathrm{\rm upper}$, $^{234}$U, $^{238}$U$_\mathrm{\rm mid}$, $^{238}$U$_\mathrm{\rm lower}$ with $^{238}$U, $^{234}$U, $^{226}$Ra, $^{210}$Pb as the parent nuclide respectively, $^{235}$U series is also classified into $^{235}$U$_\mathrm{\rm upper}$, $^{235}$U$_\mathrm{\rm lower}$ with $^{235}$U, $^{231}$Pa respectively.
Figure \ref{alpha fit} displays the likelihood-method-fitted $\alpha$-ray spectrum within the energy range of 0.50 MeV to 1.25 MeV, with black dots representing measured data and a blue solid line indicating the aggregate of simulated models. Different radioactive equilibrium groups are shown as dotted lines. Due to the small amount of $^{238}$U$_\mathrm{\rm mid}$, the green dotted line is hidden in the underflow of the spectrum. Fitting parameters were constrained to be positive. This fitting allowed us to quantify the impurities, summarized in Table \ref{fitting result}. Upper limits are indicated at the 68\% confidence level (CL). The rightmost row in the table shows the impurity content of the raw material, Gd$_{2}$O$_{3}$, as a reference value.

\begin{table}[h]
\caption{Amount of intercrystalline impurities determined by spectral fitting of $\alpha$-rays. The unit is mBq/kg and the upper limit corresponds to 68$\%$CL.}
\label{fitting result}
\centering
{
\begin{tabular}{lccc}
\hline\hline
Radioactivity&4N GAGG&High-purity GAGG&Gd$_{2}$O$_{3}$(\textcolor{black}{raw material})\\
$^{232}$Th&288.8$\pm$19.6&10.3$\pm$0.8&1.66 $\pm$ 0.41\\
$^{238}$U$_\mathrm{\rm upper}$&911.3$\pm$10.1&125.2$\pm$1.6&$<$ 16.3\\
$^{238}$U$_\mathrm{\rm mid}$&16.5$\pm$3.5&$<$0.28&$<$ 0.43\\
$^{238}$U$_\mathrm{\rm lower}$&655.5$\pm$5.8&5.93$\pm$0.44&-\\
$^{235}$U$_\mathrm{\rm upper}$&$<$22.0&4.06$\pm$1.86&$<$ 10.0\\
$^{235}$U$_\mathrm{\rm lower}$&73.5$\pm$15.3&3.07$\pm$0.68&$<$ 1.19\\
$^{234}$U&891.4$\pm$20.9&154.6$\pm$2.4&-\\\hline\hline

\end{tabular}
}
\end{table}

\begin{figure}[h]
\centering\includegraphics[width=5.0in]{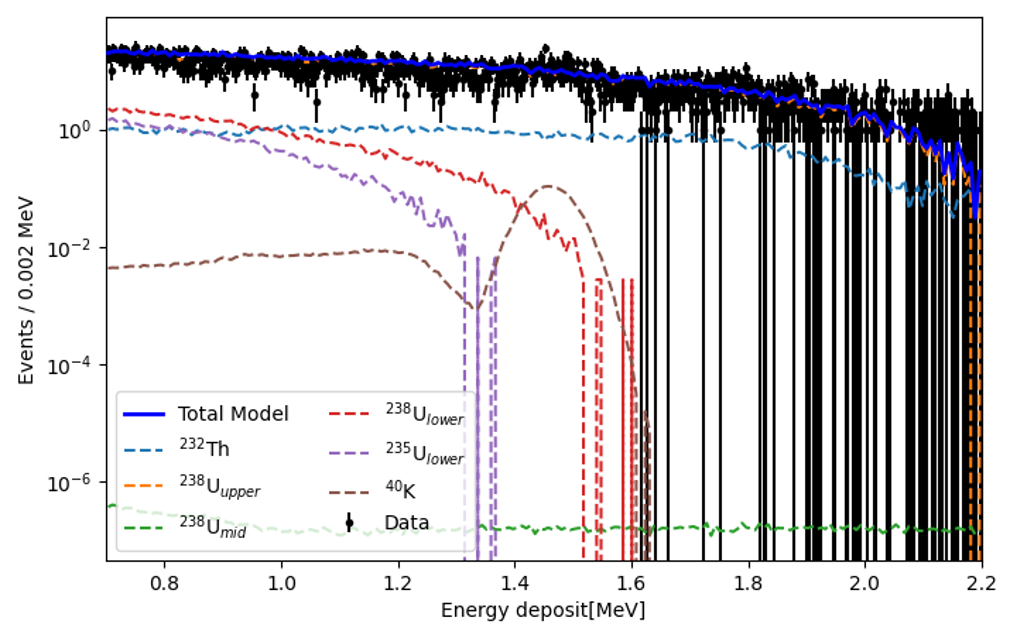}
\caption{Result of reproduction $\beta$-ray spectra of High-purity GAGG with BG model. Black points with error \textcolor{black}{bars} represent the measured $\beta$-ray background, plotted with the background model (Blue solid line), which is the total of the spectra of each radioactive equilibrium group (Dotted lines) shown in the Legend. The background models other than $^{40}$K are fixed to the value of $\alpha$-ray fitting results.}
\label{beta fit}
\end{figure}

Besides impurities in GAGG, $\beta$-ray background also originates from $\gamma$-ray emitted by $^{40}$K in the PMT. To address this, we simulated the $\gamma$-ray energy loss in GAGG due to $^{40}$K, assuming a uniform $^{40}$K distribution in the cylindrical PMT body. Like the $\alpha$-ray analysis, the actual energy resolution was applied to the simulated energy deposit data. Figure \ref{beta fit} shows the data and the GEANT4-simulated $\beta$-ray background spectrum within the 0.70 MeV to 2.20 MeV range. The impurity-derived spectra inside the crystal were determined by the amount of internal impurities estimated from the $\alpha$-rays shown in Table $\ref{fitting result}$. In addition, background spectra of $^{40}$K origin were simulated in GEANT4 and fitted additionally. Consequently, the estimated amount of $^{40}$K in background data was determined to be 0.74 $\pm$ 0.64 counts/sec/kg.

\section{Discussion}

\subsection*{Comparison of 0$\nu$2$\beta$ search sensitivity with previous study}

\textcolor{black}{
The background level of high-purity GAGG obtained in this study was used to estimate the search sensitivity of Phase 1 of the PIKACHU experiment.
The search sensitivity for the 0$\nu$2$\beta$ is expressed by the following equation.}

\begin{equation}\label{sens}
\begin{split}
{\color{black}
T^{0\nu}_{1/2} = (\rm ln \ 2) \it N_{a}\frac{\mathrm{a}}{\mathrm{A}}\epsilon\sqrt{\frac{M\cdot t}{\mathrm{BG}\cdot \Delta \mathrm{E}}}
}
\end{split}
\end{equation}

\begin{figure}[h]
\centering\includegraphics[width=5.0in]{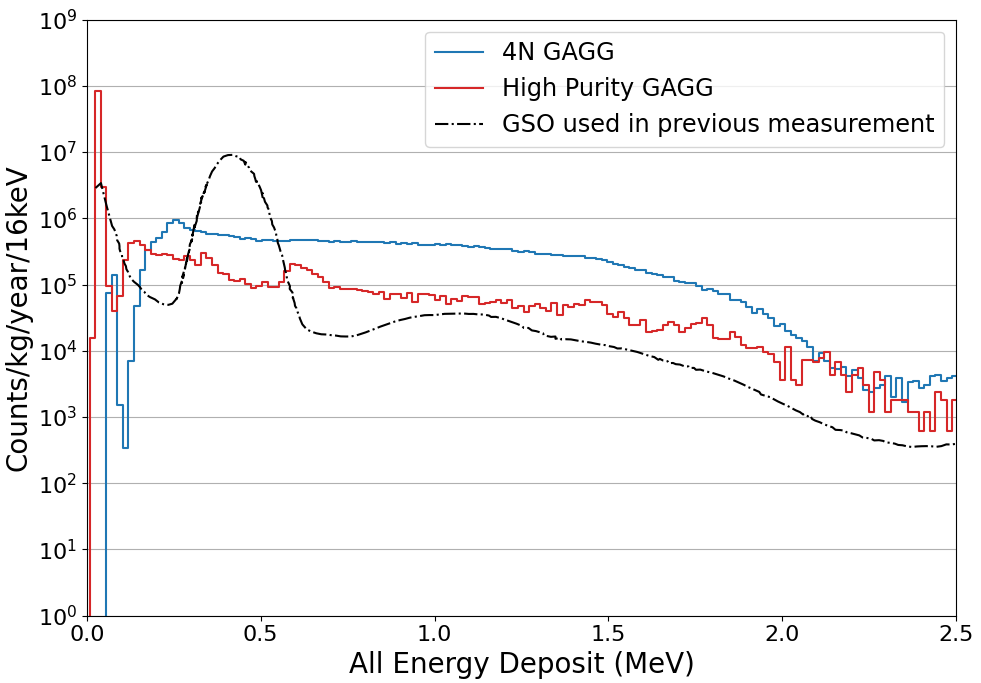}
\caption{Comparison of $\beta$-ray background spectra with 4N GAGG (blue), high-purity GAGG (red), \textcolor{black}{and $\alpha$, $\beta$-ray background spectra} with GSO from a previous study (black).}
\label{beta spectrum}
\end{figure}

\textcolor{black}{
If two High-purity GAGG crystals (diameter: 6.5 cm, length: 14.5 cm) are utilized for the same duration as in the previous study, the sensitivity calculated from Eq. (\ref{sens}) is 4.40 × 10$^{21}$ years. Here, the Avogadro number ($N_{a}$) is 6.02 × 10$^{23}$ atoms/mol, the natural abundance of $^{160}$Gd (a) is 0.219, the atomic mass of $^{160}$Gd (A) is 0.157 kg/mol, the detection efficiency ($\epsilon$) is assumed as 1.00, the detector mass ($M$) is 6.40 kg, the search period ($t$) is 1.59 years, consistent with the previous study, and the background within 1.73 ± 1.5 $\sigma$ MeV (BG $\cdot$ $\Delta$E) is 17.7 $\times$ 10$^4$ events/kg/years.
The BG $\cdot$ $\Delta$E has been computed from the number of $\beta$-ray events in the background spectrum depicted in Fig. \ref{beta spectrum}, where the black dotted line represents $\alpha$, $\beta$-ray background spectrum measured in the previous experiment (Ref. \cite{Ukraine}), and the blue and red solid lines represent the $\beta$-ray spectrum for the events selected by PSD in 4N GAGG and high-purity GAGG, respectively. 
The calculated search sensitivity of Phase 1 of the PIKACHU experiment exceeds the half-life limit of the previous study, indicating the successful development of GAGG crystals for the Phase 1 experiment.}

\textcolor{black}{To further enhance sensitivity by an order of magnitude in Phase 2, ultra-high purity crystals with background reduced by another order of magnitude need to be developed, and an experiment with 20 ultra-high purity crystals should be conducted. This would elevate the sensitivity to the same level as the expected half-life outlined in Ref. \cite{Hinohara}.}

\subsection*{Dominant background in search : $^{234m}$Pa}
\begin{figure}[h]
\centering\includegraphics[width=5.0in]{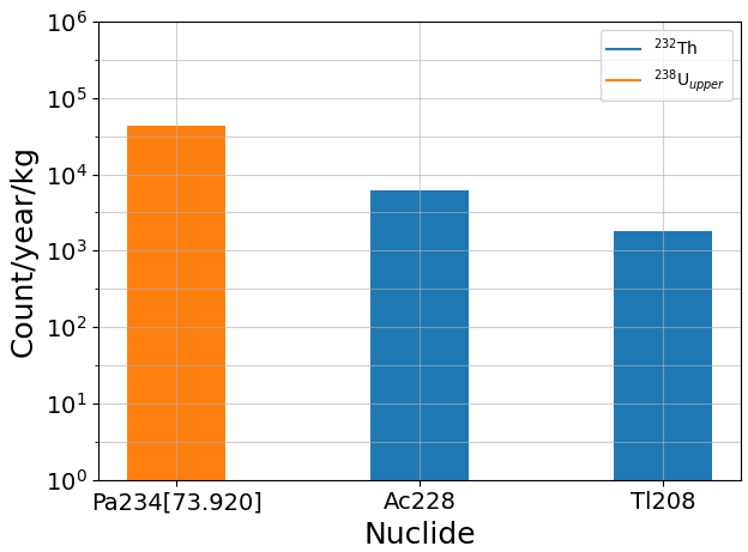}
\caption{
The amount of $\beta$-ray background within 1.73 $\pm$ 1.5 $\sigma$ MeV region, which is obtained for each parent nuclide, and the three nuclides with particularly large effects are shown in the figure.}
\label{Pa}
\end{figure}

The key issue identified is contamination of the crystals by  $^{238}$U$_\mathrm{\rm upper}$, which is about an order of magnitude higher than the other impurities (shown in Table \ref{fitting result}). This is because nuclides with $\beta$-ray $Q$-values higher than $^{160}$Gd, such as $^{234}$Pa and its metastable state $^{234m}$Pa, may contribute significantly to the background in $Q$-value region. 
 $^{234m}$Pa decays to $^{234}$U ($Q_{\beta} =$ 2.271 MeV, $T_{1/2} = $ 1.17 $min$) at a rate of 99.84$\%$ and $^{234m}$Pa undergoes isometric transition to $^{234}$Pa ($Q_{\beta} =$ 73.92 keV, $T_{1/2} =$ 1.17 $\rm min$) at the rate of 0.16$\%$. And the $^{234}$Pa decays to $^{234}$U ($Q_{\beta} =$ 2.197 MeV, $T_{1/2} =$ 6.7 $\rm hours$). Figure \ref{Pa} shows the breakdown of $\beta$-ray background nuclide in 1.73 $\pm$ 1.5 $\sigma$ MeV revealed from background fitting by GEANT4. Where Pa234[73.92] represents $^{234m}$Pa. Although $^{228}$Ac and $^{208}$Tl in $^{232}$Th decay series are also included, $^{234m}$Pa is about an order of magnitude more abundant than them and is the most dominant background in PIKACHU experiment. 

At 1.73 $\pm$ 1.5 $\sigma$ MeV, the expected background value by GEANT4 is 338.3 events, compared to 341 events in the real data. This result indicates that our background model reproduces the experimental data very well.

\subsection*{Future outlook}
The purification of GAGG's raw materials and the development of High-purity GAGG, as discussed in the paper, enable a highly sensitive search for $^{160}$Gd double beta decay. Plans are underway to use two large High-purity GAGG crystals (same size as the crystals in Fig. \ref{Fig1})  in a long-term PIKACHU experiment \textcolor{black}{Phase 1} starting in FY2024. In parallel with the long-term search, it is necessary to develop GAGG that are as free of $^{238}$U$_\mathrm{\rm upper}$ as possible; comparing Tables \ref{Raw material} and \ref{fitting result}, the crystal contains more $^{238}$U$_\mathrm{\rm upper}$ than the raw material, suggesting that the crystal were contaminated by something other than the raw material during growth. The most likely cause of radioactive contamination is insulation made from Zirconia (ZrO$_{2}$). Therefore, we are currently working on crystal growth with a structural device to prevent insulation from penetrating into the melting furnace and trying to develop GAGG crystal that are as pure as the raw material.

If the amount of $^{238}$U$_\mathrm{\rm upper}$ can be reduced by an order of magnitude compared to that in the present High-purity GAGG, we will need $^{40}$K-free PMT for 2$\nu$2$\beta$ because the photoelectric peak due to $\gamma$-rays from $^{40}$K can limit the sensitivity (Fig. \ref{beta fit}). $^{40}$K is mainly contained in the borosilicate glass used in the window of the PMT. Some underground experiments develop low radioactive PMT by converting borosilicate glass to $^{40}$K-free materials like quartz glass \cite{PMT}. We take into account the possibility of using PMT R8520-506 (Hamamatsu Photonics), with a bialkali photocathode and photosensitive window made from synthetic quartz, in PIKACHU experiments. This advancement could significantly reduce the background noise and enhance the sensitivity of the 2$\nu$2$\beta$ search.

\section*{Acknowledgment}

We thank the Research Center for Neutrino Science, Tohoku University, for allowing us to use their experimental facility for our low background experiments.
We are also grateful to Dr. Nobuo Hinohara \textcolor{black}{of the University of Tsukuba} for informative discussions on nuclear matrix element calculations, to Dr. Yoichi Tamagawa and Dr. Kyohei Nakajima \textcolor{black}{of the University of Fukui} for engaging in discussions regarding the double beta decay of $^{160}$Gd, and to Dr. Keisuke Sueki of the University of Tsukuba for help with the Ge detector measurements.
This work was supported by JSPS KAKENHI Grant No. 22H04570 and 23H01196, and the Inter-University Cooperative Research Program of the Institute for Materials Research, Tohoku University (Proposal No. 202112-RDKGE-0005 and 202012-RDKGE-0016).


%

\vspace{0.2cm}
\noindent


\let\doi\relax


\end{document}